\documentclass{article}
\usepackage{amsmath,amssymb}
\usepackage{cite}
\newcommand{\be}{\begin{equation}}
\newcommand{\ee}{\end{equation}}
\newcommand{\bea}{\begin{eqnarray}}
\newcommand{\eea}{\end{eqnarray}}
\newcommand{\bean}{\begin{eqnarray*}}
\newcommand{\eean}{\end{eqnarray*}}
\def\beq{\begin{equation}}

\def\eeq{\end{equation}}
\def\R{\mathcal{R}}

\relax

\def\d{\partial}

\def\a{\alpha'}

\begin{document}
%%%%%%%%%%%%%%%%%%%%%%%%%%%%%%%%%%%%%%%%%%%%%%%%%%%%%%%%%%%%%%%%%
%%%%%%%%%%%%%%%%%%%%%%%%%%%%%%%%%%%%%%%%%%%%%%%%%%%%%%%%%%%%%%%%%
\begin{titlepage}
\begin{center}

\vskip 20mm

{\Huge Eikonal quasinormal modes and shadow of string-corrected $d$-dimensional black holes}

\vskip 10mm

Filipe Moura$^{\dag}$ and Jo\~ao Rodrigues$^{\ddag}$

\vskip 4mm

\emph{$^{\dag}$Departamento de Matem\'atica, Escola de Tecnologias e Arquitetura and \\Instituto de Telecomunica\c c\~oes, \\ISCTE - Instituto Universit\'ario de Lisboa,
\\Av. das For\c cas Armadas, 1649-026 Lisboa, Portugal}
\vskip 2 mm
\emph{$^{\ddag}$Centro de F\'\i sica da Universidade de Coimbra,\\ Rua Larga, 3004-516 Coimbra, Portugal}

\vskip 4mm
\texttt{fmoura@lx.it.pt}, \quad
\texttt{jvrodrigues@pollux.fis.uc.pt}

\vskip 6mm

\end{center}

\vskip .2in

\begin{center} {\bf Abstract } \end{center}
\begin{quotation}\noindent
We compute the quasinormal frequencies of $d$-dimensional spherically symmetric black holes with leading string $\a$ corrections in the eikonal limit for tensorial gravitational perturbations and scalar test fields. We find that, differently than in Einstein gravity, the real parts of the frequency are no longer equal for these two cases. The corresponding imaginary parts remain equal to the principal Lyapunov exponent corresponding to circular null geodesics, to first order in $\a$. We also compute the radius of the shadow cast by these black holes.
\end{quotation}

\vfill

%%%%%%%%%%%%%%%%%%

\end{titlepage}

\eject

\newpage

%%%%%%%%%%%%%%%%%%%%%%%%%%%%%%%%%%%%%%%%%%%%%%%%%%%%%%%%%%%%%%%%%%%%%%
\section{Introduction}
%%%%%%%%%%%%%%%%%%%%%%%%%%%%%%%%%%%%%%%%%%%%%%%%%%%%%%%%%%%%%%%%%%%%%%
\noindent

Black holes in the ringdown phase resulting from a black hole collision form a dissipative system which can be described by black hole perturbation theory: they lose energy by emitting gravitational radiation. Those decaying oscillations are called quasinormal modes, and they are given in terms of complex frequencies. The spectrum of black hole quasinormal modes (QNMs) does not depend on what drives the perturbations: it is completely determined in terms of intrinsic physical quantities of the black hole such as mass, charge or spin, and eventually (beyond Einstein gravity) some other parameters of the theory. This feature turns QNMs into preferential probes for testing theories of gravity beyond Einstein, since the ringing frequencies can be directly measured by the gravitational wave detectors. In spite of their classical origin, QNMs may provide a glimpse into the eventual quantum nature of the black hole. With the advent of gravitational wave astronomy, therefore, interest in the study of black hole QNMs has raised. Some recent works computing QNMs of black holes with higher derivative corrections coming from different theories are \cite{Blazquez-Salcedo:2016enn, Chen:2018vuw, Chen:2019dip, Cano:2020cao, Konoplya:2017wot, Konoplya:2020bxa, Pierini:2021jxd, Moura:2021nuh}.

Most of the times, QNMs have to be computed numerically. Nonetheless, different analytical methods have been developed in order to compute QNMs in some limiting cases. One of such cases is the eikonal limit. Experimental data related to black holes coming either from the detection of gravitational waves or from observation of the electromagnetic spectrum in their vicinity, does not fix the near-horizon geometry; more information for that purpose can be obtained from the shadow cast by a black hole. Some recent studies of black hole shadows in theories beyond Einstein gravity are \cite{Banerjee:2019nnj, Khodadi:2020jij, Campos:2021sff}. For asymptotically flat spherically symmetric black holes, under some conditions that we discuss in section \ref{shadow}, the radius of such shadow can be related with the real part of QNM frequencies in the eikonal limit, as we will see.

In this article we will consider $d$-dimensional asymptotically flat spherically symmetric black holes in string theory and compute their quasinormal frequencies in the eikonal limit and the radius of the black hole shadow. The article is organized as follows: in section \ref{lg} we will review the string-corrected black hole solution we will consider, the master equation and the corresponding potentials for minimally coupled test scalar fields and tensorial gravitational perturbations; in section \ref{eik} we analyze how to compute the corresponding quasinormal frequencies in the eikonal limit in this context, and we obtain such frequencies; in section \ref{shadow} we review how to obtain the black hole shadow from the quasinormal modes and we compute it for the black hole we consider.

%%%%%%%%%%%%%%%%%%%%%%%%%%%%%%%%%%%%%%%%%%%%%%%%%%%%%%%%%%%%%%%%%%%%%%
\section{String-corrected spherically symmetric black holes and their tensorial perturbations}
%%%%%%%%%%%%%%%%%%%%%%%%%%%%%%%%%%%%%%%%%%%%%%%%%%%%%%%%%%%%%%%%%%%%%%
\label{lg}
\noindent

A general static spherically symmetric metric in $d$ dimensions can always be cast in the form
\be \label{schwarz}
ds^2 = -f(r)\ dt^2  + f^{-1}(r)\ dr^2 + r^2 d\Omega^2_{d-2}.
\ee
A minimally coupled scalar field propagating in the background of a black hole of the form (\ref{schwarz}) can be expanded as
\begin{equation}
\Phi(t,r,\theta)= e^{i\omega t} \sum_{\ell} \Phi_{\ell, \omega}(r) Y_{\ell}(\theta)\,,
\label{sphericalharmonics1}
\end{equation}
where $\omega$ is the wave frequency, $\ell$ is the angular quantum number associated with the polar angle $\theta$ and $Y_{\ell} (\theta)$ are the usual spherical harmonics defined over the $(d-2)$ unit sphere $\mathbb{S}^{d-2}$, with the azimuthal angles in this case set to constants. In terms of the tortoise coordinate $x$ defined by
\be
dx = \frac{dr}{f(r)}, \label{tort}
\ee
the field equation satisfied by each component $\Phi_{\ell, \omega}(r)$ (which we denote by $\psi$) is given by
\be
\frac{d^2 \psi}{d\, x^2} + \omega^2 \psi = V \left[ f(r) \right] \psi \label{potential0}
\ee
with the potential \cite{hep-th/0206084}
\be
V_{\textsf{0}} [f(r)] = f(r) \left( \frac{\ell \left( \ell + d - 3 \right)}{r^2} + \frac{\left( d - 2 \right) \left( d - 4 \right) f(r)}{4r^2} + \frac{\left( d - 2 \right) f'(r)}{2r} \right). \label{v0}
\ee

Perturbations of a $d-$dimensional spherically symmetric metric can be decomposed according to their tensorial behavior on $\mathbb{S}^{d-2}$ as scalar, vectorial or (for $d>4$) tensorial. Each type of perturbation is described in terms of a master variable, which we also denote generically by $\psi$. In Einstein gravity, this master variable obeys a second order differential equation (``master equation'') analogous to (\ref{potential0}), but with a potential that also depends on the kind of perturbation one considers \cite{ik03a}. Specifically, tensorial perturbations obey an equation (\ref{potential0}) with the same potential (\ref{v0}) of scalar test fields.

In the presence of higher order corrections in the lagrangian, one can still have spherically symmetric black holes of the form (\ref{schwarz}), but the master equation obeyed by each perturbation variable is expected to change. %The master equations for tensorial, vectorial and scalar perturbations of a metric of the form (\ref{schwarz}) in the presence of second order Gauss-Bonnet corrections, and the respective potentials, have been worked out in \cite{Dotti:2005sq,Gleiser:2005ra}. In all these cases a second order master equation was obtained.
Tensorial metric perturbations in the presence of leading $\a$ corrections have been studied in \cite{Moura:2006pz,Moura:2012fq}, $\a$ being the inverse string tension. Concretely, the following $d$--dimensional effective action with purely gravitational $\a$ corrections has been considered:
\be \label{eef} \frac{1}{16 \pi G} \int \sqrt{-g} \left( \R -
\frac{4}{d-2} \left( \d^\mu \phi \right) \d_\mu \phi +
\mbox{e}^{\frac{4}{d-2} \phi} \frac{\lambda}{2}\
\R^{\mu\nu\rho\sigma} \R_{\mu\nu\rho\sigma} \right) \mbox{d}^dx .
\ee
This is the effective action of bosonic and heterotic string theories, to first order in $\a$, with $\lambda = \frac{\a}{2}, \frac{\a}{4}$, respectively. \footnote{Type II superstring theories do not have $\a$ corrections to this order.} In both cases one can consistently set all other bosonic and fermionic fields present in the string spectrum to zero except for the dilaton field $\phi$. We are also not considering compactification effects: action (\ref{eef}) results from compactifying on a flat torus, leaving $d$ dimensions uncompactified.

In \cite{Moura:2006pz,Moura:2012fq} it has been shown that, perturbing the field equations resulting from this action, for tensorial perturbations of a spherically symmetric metric (\ref{schwarz}) one also obtains a second order master equation (\ref{potential0}). The corresponding potential is given by
\bea
V_{\textsf{T}} [f(r)] &=& \lambda\ \frac{f(r)}{r^2} \left[ \left( \frac{2 \ell \left( \ell + d - 3 \right)}{r} + \frac{\left( d - 4 \right) \left( d - 5 \right) f(r)}{r} + \left( d - 4 \right) f'(r) \right) \left( 2 \frac{1 - f(r)}{r} + f'(r) \right) \right. \nonumber \\
&+& \left. \Big( 4 (d-3) - (5d-16) f(r) \Big) \frac{f'(r)}{r} - 4 \left( f'(r) \right)^2 + \left( d-4 \right) f(r) f''(r) \right] + V_{\textsf{0}} [f(r)], \label{vt}
\eea
with $V_{\textsf{0}} [f(r)]$ given by (\ref{v0}).

Spherically symmetric $d-$dimensional black hole solutions with leading $\a$ corrections have been obtained in \cite{cmp89,Moura:2009it}. Specifically concerning the action (\ref{eef}), a solution of the respective field equations is of the form (\ref{schwarz}), with

\bea \label{fr2}
f(r) &=& f_0(r) \left(1+ \frac{\lambda}{R_H^2} \delta f(r) \right), \\
f_0(r) &=& 1 - \frac{R_H^{d-3}}{r^{d-3}}, \label{fr0}\\
\delta f(r) &=& - \frac{(d-3)(d-4)}{2}\ \frac{R^{d-3}_H}{r^{d-3}}\ \frac{1 - \frac{R_H^{d-1}}{r^{d-1}}}{1 - \frac{R^{d-3}_H}{r^{d-3}}}.
\eea
When $\lambda=0$ we get the Tangherlini solution with $f(r)=f_0(r)$. This is then clearly an $\a-$corrected Tangherlini solution, which has been obtained and studied by Callan, Myers and Perry in \cite{cmp89}. Its only horizon occurs at the same radius $r=R_H$ of the Tangherlini solution. For our purposes, it is enough to just quote here the explicit expression for the temperature of this black hole, given by
\be
T = \frac{d-3}{4 \pi R_H} \left( 1 - \frac{\left( d-1 \right) \left( d-4 \right)}{2}\ \frac{\lambda}{R_H^2} \right). \label{temp}
\ee

%%%%%%%%%%%%%%%%%%%%%%%%%%%%%%%%%%%%%%%%%%%%%%%%%%%%%%%%%%%%%%%%%
%%%%%%%%%%%%%%%%%%%%%%%%%%%%%%%%%%%%%%%%%%%%%%%%%%%%%%%%%%%%%%%%%
\section{Quasinormal modes in the eikonal limit}
%%%%%%%%%%%%%%%%%%%%%%%%%%%%%%%%%%%%%%%%%%%%%%%%%%%%%%%%%%%%%%%%%
%%%%%%%%%%%%%%%%%%%%%%%%%%%%%%%%%%%%%%%%%%%%%%%%%%%%%%%%%%%%%%%%%
\label{eik}
\noindent

In this section, we compute quasinormal modes in the eikonal limit for the cases we have been considering: minimally coupled (test) scalar fields and tensorial gravitational perturbations. The eikonal limit of quasinormal modes corresponds to having a large multipole number $\ell$, or equivalently a large real part of the QNM frequency. In this limit one can apply the WKB approximation.

In the eikonal limit ($\ell \rightarrow +\infty$), we are left with only one term from the potential (\ref{v0}):
\be
V_0^{\textsf{eik}} [f(r)] = \ell^2 \frac{f(r)}{r^2}. \label{v0eik}
\ee
In Einstein gravity (i.e. taking $\lambda=0$), this limit is the same for the potentials corresponding not only to test scalar fields but also to all kinds of gravitational perturbations: tensorial, vectorial and scalar \cite{ik03a}, and so are the corresponding eikonal quasinormal modes. We now briefly review how to compute these quasinormal modes.

In order to apply the WKB approximation, we assume that the potential we are dealing with vanishes at the horizon and at infinity, and that it has a single maximum for some value $r > R_H$. Concretely for $V_0^{\textsf{eik}} [f(r)]$, such maximization corresponds to having
\be
2 f(r) = r f'(r), \label{circ}
\ee
which is precisely the equation defining a circular null geodesic. Indeed, in \cite{Cardoso:2008bp} it was shown that for $d$-dimensional asymptotically flat spherically symmetric black holes in Einstein gravity the eikonal quasinormal frequencies are given in terms of parameters of the unstable circular null geodesics around such black holes.
Concretely, the quasinormal frequency corresponding to the overtone mode $n$ is given by
\be
\omega_n = \ell \Omega -i \left(n + \frac{1}{2} \right) \Lambda, \label{omegaeik}
\ee
where $\Omega$ is the angular velocity at such geodesics and $\Lambda$ is the principal Lyapunov exponent corresponding to such orbit. These quantities are given by
\bea
\Omega&=&\frac{\sqrt{f(r_c)}}{r_c}, \label{omegageo} \\
\Lambda &=& \sqrt{-\frac{r_c^2}{2 f(r_c)}\frac{d^2}{d x^2}  \left(\frac{f}{r^2}\right)_{r=r_c}} \nonumber \\
&=&  \sqrt{-\frac{r_c^2}{2} \left(f' \left(\frac{f}{r^2}\right)'+ f \left(\frac{f}{r^2}\right)''\right)_{r=r_c}}, \label{lyap}
\eea
with $r_c$ being the value of $r$ maximizing $V_0^{\textsf{eik}} [f(r)]$, i.e. the radius of the circular null geodesic, a solution of (\ref{circ}). $x$ is the tortoise coordinate (\ref{tort}).

In Einstein gravity we have the Tangherlini solution with $f(r)=f_0(r)$, as we saw. The corresponding solution to (\ref{circ}) is given by
\be
r_0=R_H \sqrt[d-3]{\frac{d-1}{2}}. \label{r0}
\ee
The quasinormal frequencies are obtained by simply replacing $r_c=r_0$ in (\ref{omegageo}) and (\ref{lyap}).

We wish to generalize the study of quasinormal modes in the eikonal limit for the string--corrected Callan-Myers-Perry black hole given by (\ref{fr2}). For simplicity, we start by QNMs of test scalar fields, after which we study QNMs of tensorial gravitational perturbations.

%%%%%%%%%%%%%%%%%%%%%%%%%%%%%%%%%%%%%%%%%%%%%%%%%%%%%%%%%%%%%%%%%%%%%%
\subsection{The eikonal limit for test scalar fields}
%%%%%%%%%%%%%%%%%%%%%%%%%%%%%%%%%%%%%%%%%%%%%%%%%%%%%%%%%%%%%%%%%%%%%%
\noindent

The procedure for obtaining QNMs in the eikonal limit for test scalar fields is essentially identical to the one we just saw in Einstein gravity. We want to maximize the same potential, given by (\ref{v0eik}). Such maximization corresponds to solving equation (\ref{circ}). The only change comes from the function $f(r)$, which is now given by (\ref{fr2}), including a $\lambda$ correction, corresponding to the solution we have been considering. For such choice of $f(r)$, the value of $r$ which is a solution of (\ref{circ}) is given, to first order in $\lambda$, by
\be
r_c=R_H \sqrt[d-3]{\frac{d-1}{2}}\left(1+\frac{\lambda}{R_H^2}\left(\frac{d-4}{2}-(d-4) \left(\frac{2}{d-1}\right)^{\frac{d-1}{d-3}}\right)\right). \label{rct}
\ee
Replacing this value in (\ref{omegageo}) and (\ref{lyap}), respectively, we obtain
\bea
\Omega &=& \sqrt{\frac{d-3}{2}} \frac{1}{R_H} \left(\frac{2}{d - 1}\right)^{\frac{d-1}{2(d-3)}} \left[1 + \frac{\lambda}{R_H^2}\frac{d-4}{2} \left( \left(\frac{2}{d - 1}\right)^{\frac{d-1}{d-3}} -1 \right) \right], \label{omegat} \\
\Lambda &=&\frac{d-3}{\sqrt{2}} \left(\frac{2}{d - 1}\right)^{\frac{d-1}{2(d-3)}} \frac{1}{R_H} \left[1 - \frac{\lambda}{R_H^2}\frac{d-4}{2} \left(
1+ (d-2) \left(\frac{2}{d - 1}\right)^{\frac{d-1}{d-3}} \right) \right] \label{lyapt}.
\eea
The $\lambda=0$ limits of (\ref{rct}), (\ref{omegat}) and (\ref{lyapt}) correspond to the known results in Einstein gravity.

We now express the results in terms of the black hole temperature. We simply invert, to first order in $\lambda$, the expression $T(R_H)$ given by (\ref{temp}); we then replace $R_H$ by the result $R_H(T)$ in (\ref{omegat}) and (\ref{lyapt}), obtaining
\bea
\Omega &=& 2 \pi T \sqrt{\frac{2}{d-3}} \left(\frac{2}{d - 1}\right)^{\frac{d-1}{2(d-3)}} \left[1 + 8 \pi^2 T^2 \lambda \frac{d-4}{(d-3)^2} \left( d - 2 + \left(\frac{2}{d - 1}\right)^{\frac{d-1}{d-3}} \right) \right]; \\
\Lambda &=& \frac{4 \pi T}{\sqrt{2}} \left(\frac{2}{d - 1}\right)^{\frac{d-1}{2(d-3)}} \left[1 + 8 \pi^2 T^2 \lambda \frac{(d-2)(d-4)}{(d-3)^2} \left( 1-\left(\frac{2}{d - 1}\right)^{\frac{d-1}{d-3}} \right) \right].
\eea
We can check that, when expressed in terms of a physical variable (in this case the temperature), $\a$ corrections are positive for both real and imaginary parts of the QNM frequencies.

%%%%%%%%%%%%%%%%%%%%%%%%%%%%%%%%%%%%%%%%%%%%%%%%%%%%%%%%%%%%%%%%%%%%%%
\subsection{The eikonal limit for tensorial perturbations}
%%%%%%%%%%%%%%%%%%%%%%%%%%%%%%%%%%%%%%%%%%%%%%%%%%%%%%%%%%%%%%%%%%%%%%
\noindent

We now turn to the computation of QNM frequencies in the eikonal limit for gravitational tensorial perturbations, described to first order in $\a$ by the potential (\ref{vt}). In order to distinguish from the scalar field case we have just seen, we denote these frequencies by
\be
\omega^{\textsf{T}}_n = \ell \Omega_{\textsf{T}} -i \left(n + \frac{1}{2} \right) \Lambda_{\textsf{T}}. \label{omegaeikt}
\ee
The eikonal limit ($\ell \rightarrow +\infty$) of the potential (\ref{vt}), compared to the one of (\ref{v0}), naturally has a $\lambda$ correction. This limit can be expressed analogously to (\ref{v0eik}), but with $f(r)$ replaced by $g(r)$:
\bea
V_{\textsf{T}}^{\textsf{eik}} [f(r)] &=& \ell^2 \frac{g(r)}{r^2}, \label{potg} \\
g(r) &\equiv& f(r)\left( 1 + \frac{2 \lambda}{r^2} \left( 2 \left(1 - f(r) \right)+ r f'(r) \right) \right). \label{gr}
\eea
Because of the $\lambda$ correction in $g(r)$, the equation for the maximization of $V_{\textsf{T}}^{\textsf{eik}}$ is no longer the equation defining a circular null geodesic, as pointed out in \cite{Konoplya:2017wot}; in our case, it is rather given by
\be
2 g(r) = r g'(r). \label{circt}
\ee
Its solution is given, to first order in $\lambda$, by
\be  \label{rcp}
r_t=R_H \sqrt[d-3]{\frac{d-1}{2}}\left(1+\frac{\lambda}{R_H^2}\left(\frac{d-4}{2}-(2d-5) \left(\frac{2}{d-1}\right)^{\frac{d-1}{d-3}}\right)\right).
\ee
The procedure for the computation of $\Omega_{\textsf{T}},\Lambda_{\textsf{T}}$ in (\ref{omegaeikt}) is similar to the one described in the previous section for $\Omega,\Lambda$. We simply have to take $g(r)$ instead of $f(r)$ and $r_t$ instead of $r_c$ in (\ref{omegageo}) and (\ref{lyap}):
\bea
\Omega_{\textsf{T}}&=&\frac{\sqrt{g(r_t)}}{r_t}, \\
\Lambda_{\textsf{T}} &=& \sqrt{-\frac{r_t^2}{2 g(r_t)}\frac{d^2}{d x^2}  \left(\frac{g}{r^2}\right)_{r=r_t}} \nonumber \\
&=&  \sqrt{-\frac{r_t^2}{2} \frac{f(r_t)}{g(r_t)} \left(f' \left(\frac{g}{r^2}\right)'+ f \left(\frac{g}{r^2}\right)''\right)_{r=r_t}}. \label{lyptt}
\eea
We obtained the following results, to first order in $\lambda$:
\bea
\Omega_{\textsf{T}} &=& \sqrt{\frac{d-3}{2}} \frac{1}{R_H} \left(\frac{2}{d - 1}\right)^{\frac{d-1}{2(d-3)}} \left[1 + \frac{\lambda}{R_H^2}\frac{1}{2} \left(3 (d-2) \left(\frac{2}{d - 1}\right)^{\frac{d-1}{d-3}} -(d-4) \right) \right], \label{omegap} \\
\Lambda_{\textsf{T}} &=&(d-3) \left(\frac{2}{d - 1}\right)^{\frac{d-1}{2(d-3)}} \frac{1}{R_H} \left[1 - \frac{\lambda}{R_H^2}\frac{d-4}{2} \left(
1+ (d-2) \left(\frac{2}{d - 1}\right)^{\frac{d-1}{d-3}} \right) \right]. \label{lyapp}
\eea
The $\lambda=0$ limits of (\ref{rcp}), (\ref{omegap}) and (\ref{lyapp}) correspond to the known results in Einstein gravity.

Like we did for the case of test scalar fields, we now express the results in terms of the black hole temperature, given by (\ref{temp}):
\bea
\Omega_{\textsf{T}} &=& 2 \pi T \sqrt{\frac{2}{d-3}} \left(\frac{2}{d - 1}\right)^{\frac{d-1}{2(d-3)}} \left[1 + 8 \pi^2 T^2 \lambda \frac{d-2}{(d-3)^2} \left( d - 4 + 3\left(\frac{2}{d - 1}\right)^{\frac{d-1}{d-3}} \right) \right], \\
\Lambda_{\textsf{T}} &=& 4 \pi T \left(\frac{2}{d - 1}\right)^{\frac{d-1}{2(d-3)}} \left[1 + 8 \pi^2 T^2 \lambda \frac{(d-2)(d-4)}{(d-3)^2} \left( 1-\left(\frac{2}{d - 1}\right)^{\frac{d-1}{d-3}} \right) \right].
\eea
Like for the case of test scalar fields these $\a$ corrections, expressed in terms of the temperature, are positive both for the real and for the imaginary parts of the QNM frequencies.

%%%%%%%%%%%%%%%%%%%%%%%%%%%%%%%%%%%%%%%%%%%%%%%%%%%%%%%%%%%%%%%%%
%%%%%%%%%%%%%%%%%%%%%%%%%%%%%%%%%%%%%%%%%%%%%%%%%%%%%%%%%%%%%%%%%
\subsection{The equality of tensorial and test imaginary parts}
%%%%%%%%%%%%%%%%%%%%%%%%%%%%%%%%%%%%%%%%%%%%%%%%%%%%%%%%%%%%%%%%%
%%%%%%%%%%%%%%%%%%%%%%%%%%%%%%%%%%%%%%%%%%%%%%%%%%%%%%%%%%%%%%%%%
\label{equality}
\noindent

We must point out that, because the equation for the maximization of $V_{\textsf{T}}^{\textsf{eik}}$ does not coincide with the one defining a circular null geodesic, the identifications of $\Omega_{\textsf{T}}$ with the angular velocity at such geodesic and $\Lambda_{\textsf{T}}$ as the principal Lyapunov exponent corresponding to such orbit are not expected to be valid for these perturbations. That is the case for $\Omega_{\textsf{T}}$: by comparing (\ref{omegat}) to (\ref{omegap}), we see that $\Omega$ and $\Omega_{\textsf{T}}$ are indeed different. But that is not the case for $\Lambda_{\textsf{T}}$.

By comparing (\ref{lyapt}) to (\ref{lyapp}), we notice that the imaginary components of the eikonal quasinormal frequencies, associated with scalar test fields and tensor type gravitational perturbations, are equal \emph{to first order in $\lambda$}:
\be
\Lambda=\Lambda_{\textsf{T}} + \mathcal{O}(\lambda^2). \label{llt}
\ee
This property is surprising and remarkable. We saw that, in the absence of higher order corrections, quasinormal modes of tensorial perturbations and test fields were identical; in the eikonal limit, test fields and all kinds of gravitational perturbations had the same QNM spectra. In all these cases, the equality of the QNM spectra had its origin in the equality of the corresponding perturbation potentials in the eikonal limit: the same potential naturally gave rise to the same QNM frequencies. In the presence of leading string corrections, in the cases we considered\footnote{The  potentials for black holes in the presence of string $\a$ corrections are known only in the cases we considered: tensorial gravitational perturbations and test fields.} the perturbation potentials are different but the imaginary parts of QNM frequencies remain the same (not the real parts, though).

We now provide an analytical proof of this remarkable equality, under the assumptions we have been considering in this article. In order to simplify the notation for this section, we define $\lambda'=\frac{\lambda}{R_H^2}.$ We will always discard contributions of order $\mathcal{O}(\lambda'^2)$; every equation in this section should thus have an extra ``$+ \mathcal{O}(\lambda'^2)$'' term, which we omit also to simplify the notation (but it is important to keep this in mind).

We recall that $r_c$ and $r_t$ are the solutions to (\ref{circ}) and (\ref{circt}), respectively:
\bea
    2f(r_c) = r_cf'(r_c), \label{flrc} \\
    2g(r_t) = r_tg'(r_t). \label{flrt}
\eea
Here $f(r), \, g(r)$ correspond to the leading terms of the potentials $V_0^{\textsf{eik}} [f(r)], \, V_{\textsf{T}}^{\textsf{eik}} [f(r)]$. We will assume for a moment that $f(r), \, g(r)$ are generic functions, defined only by conditions (\ref{v0eik}) and (\ref{potg}) respectively, in the presence of higher order corrections proportional to a constant $\lambda'\frac{}{}$, which for a while we also take as generic. The only restriction we will consider is that, in the absence of the higher order corrections, both $f(r), \, g(r)$ reduce to $f_0(r)$ given by (\ref{fr0}). This means that both  these functions can be written in the form
\bea
    f(r) = f_0(r) + \lambda' \delta f, \nonumber \\
    g(r) = f_0(r) + \lambda' \delta g. \label{fgg}
\eea
Similarly, $r_t,\,r_c$ can also both be written in the form
\bea
    r_t = r_0 + \lambda' \delta r_t, \nonumber \\
    r_c = r_0 + \lambda' \delta r_c. \label{rcrt}
\eea
$r_0$, given by (\ref{r0}), is the solution of
\begin{equation}
    2f_0(r_0) = r_0f_0'(r_0).
    \label{195}
\end{equation}

From (\ref{lyap}) and (\ref{lyptt}), and using  (\ref{flrc}) and  (\ref{flrt}), proving (\ref{llt}) amounts to prove that
\begin{equation}
    \frac{f^2(r_c)}{r_c^2}\left[2 - \frac{f''(r_c)}{f(r_c)}r_c^2\right] = \frac{f^2(r_t)}{r_t^2}\left[2 - \frac{g''(r_t)}{g(r_t)}r_t^2\right].
    \label{199}
\end{equation}

We can define the function
\begin{equation}
    \Delta f(r) := g(r) - f(r).
    \label{delt}
\end{equation}
From (\ref{fgg}) we see that $\Delta f(r)$ is necessarily of order $\lambda'$. Similarly from (\ref{rcrt}) we can conclude that $r_c-r_t$ is also of
order $\lambda'$. Knowing these facts, we may consider first order in $\lambda'$ Taylor expansions of $f(r_c), f''(r_c)$ around $r_t$. Using such
expansions and (\ref{fgg}) we can write, always to first order in $\lambda'$,
\bea
    f(r_c)f''(r_c) &=& \left[f(r_t) + f'(r_t)(r_c-r_t)\right]\left[f''(r_t) + f'''(r_t) (r_c-r_t)\right] \nonumber \\
    &=& f(r_t)f''(r_t) + (r_c-r_t)\left[f_0(r_0)f_0'''(r_0) + f_0'(r_0)f_0''(r_0)\right] \label{ff2}
\eea
and
\begin{equation}
\begin{split}
    \frac{f^2(r_t)}{g(r_t)}g''(r_t) = f^2(r_t)\left(\frac{1}{f(r_t)} - \frac{\Delta f (r_t)}{f^2(r_t)}\right)g''(r_t)  = \left(f(r_t) - \Delta f(r_t)\right)\left(f''(r_t) + \Delta f''(r_t)\right) \\ = f(r_t)f''(r_t) - \Delta f (r_0) f''_0(r_0) + f_0(r_0)\Delta f''(r_0).
    \label{197}
    \end{split}
\end{equation}
Knowing that $\frac{1}{r_c^2}-\frac{1}{r_t^2} = -\frac{2}{r_0^3}(r_c-r_t)$, always to first order in $\lambda'$, and also using (\ref{195}), we have
\begin{equation}
    2\frac{f^2(r_c)}{r_c^2} - 2 \frac{f^2(r_t)}{r_t^2} = 2f^2(r_t)\left(\frac{1}{r_c^2}-\frac{1}{r_t^2}\right) + \frac{4}{r_0^2}f_0(r_0)f_0'(r_0)(r_c-r_t) = (r_c-r_t) \frac{4}{r_0^3}f_0^2(r_0).
    \label{194}
\end{equation}
From (\ref{ff2}), (\ref{197}) and (\ref{194}) we can rewrite equation (\ref{199}) as
\begin{equation}
(r_c-r_t) \left[ \frac{4}{r_0^3}f_0^2(r_0) - f_0(r_0)f_0'''(r_0) - f_0'(r_0)f_0''(r_0)\right] + f_0(r_0)\Delta f''(r_0)- \Delta f(r_0)f''_0(r_0) = 0.
   \label{204}
\end{equation}

Up to now, in this section we have kept the functions $f(r), g(r)$ generic. The only assumption made was about $f_0$, given by (\ref{fr0}), i.e., we are considering a solution which is a higher order correction to the Tangherlini black hole. The most general condition that such solution must follow in order to verify the equality (\ref{llt}) is given by (\ref{204}).

We now verify that (\ref{204}) is indeed fulfilled by the solution and potential for tensor-type perturbations we considered in this article, with $f(r), \, g(r)$ given respectively by (\ref{fr2}) and (\ref{gr}). Taking these functions, the values of $r_c, \,r_t$ are those given in (\ref{rct}) and (\ref{rcp}), from which we see that
\be
r_c-r_t = 2 \lambda' \frac{R_H^2}{r_0}. \label{rc-rt}
\ee
From (\ref{gr}) we may rewrite the definition (\ref{delt}) as
\begin{equation}
    \Delta f(r) =2\lambda'f(r) \left(\frac{R_H}{r}\right)^2\left[2(1-f(r))+rf'(r)\right]. \label{202}
\end{equation}
Differentiating twice $\Delta f(r)$ and also using (\ref{195}), after some algebra we get
\begin{equation}
\Delta f''(r_0)  = 4 \lambda' R_H^2\left[\frac{f_0''(r_0)}{r_0^2} - 2\frac{f_0(r_0)}{r_0^4} +\frac{1}{2}\frac{f_0(r_0)}{r_0}f_0'''(r_0)\right]
\end{equation}
which, combined with (\ref{202}), gives
\begin{equation}
f_0(r_0)\Delta f''(r_0) - \Delta f(r_0)f_0''(r_0) = \lambda' R_H^2\left[2 \frac{f_0^2(r_0)}{r_0}f_0'''(r_0) - 8 \frac{f_0^2(r_0)}{r_0^4}\right].
\end{equation}
Using (\ref{rc-rt}) and plugging the equation above in (\ref{204}) yields
\begin{equation}
\lambda' \frac{2 R_H^2}{r_0} \left[ f_0^2(r_0) f'''_0(r_0) - f_0(r_0)f_0'''(r_0)-f_0'(r_0)f_0''(r_0)\right] = 0.
\label{205}
\end{equation}
We found a condition which is equivalent to (\ref{199}), but much simpler! Considering (\ref{195}), equation (\ref{205}) holds true if and only if $r_0$ is a root of
\begin{equation}
\mathcal{F}(r) :=  f_0'''(r)\left(f_0(r)-1\right)- \frac{2}{r}f_0''(r).
\end{equation}
Using (\ref{fr2}), we can explicitly write
\begin{equation}
    \mathcal{F}(r) = \left(\frac{R_H}{r}\right)^{d-3}(d-3)(d-2)\left[\frac{2}{r^3}-(d-1)\frac{R_H^{d-3}}{r^d}\right].
\end{equation}
It is easy to see that the only real positive root of $\mathcal{F}$ is
\begin{equation}
r = \left(\frac{d-1}{2}\right)^{\frac{1}{d-3}} R_H = r_0.
\end{equation}
This completes the proof of the equality (\ref{llt}) between the imaginary components of test and tensor gravitational eikonal quasinormal frequencies of solution (\ref{fr2}) to first order in $\lambda'$, as we wanted.

%%%%%%%%%%%%%%%%%%%%%%%%%%%%%%%%%%%%%%%%%%%%%%%%%%%%%%%%%%%%%%%%%
%%%%%%%%%%%%%%%%%%%%%%%%%%%%%%%%%%%%%%%%%%%%%%%%%%%%%%%%%%%%%%%%%
\section{Black hole shadow and quasinormal modes}
%%%%%%%%%%%%%%%%%%%%%%%%%%%%%%%%%%%%%%%%%%%%%%%%%%%%%%%%%%%%%%%%%
%%%%%%%%%%%%%%%%%%%%%%%%%%%%%%%%%%%%%%%%%%%%%%%%%%%%%%%%%%%%%%%%%
\label{shadow}
\noindent

In this section, we provide an analytical expression for the radius of the shadow cast by the Callan-Myers-Perry black hole (\ref{fr2}).

In a $d$-dimensional asymptotically flat black hole space time, an observer at spatial infinity can set up an Euclidean spatial coordinate system. In
this framework, the $(d-2)$-dimensional hyperplane that simultaneously contains the black hole and is normal to the line connecting it to the observer is called the observer's sky. Now, we consider a light source behind the black hole, from the observer's perspective. As the photons, emitted by the source, travel through space time, some will get trapped by the black hole, while some will eventually reach the observer. The absence of photons that ended trapped by the black hole will make the observer perceive some sort of silhouette in the observer's sky. The shape of the corresponding shadow is determined by the impact parameters of such photons, following a null geodesic. For the case of a spherical symmetric black hole like (\ref{schwarz}), this silhouette forms a disk and the radius of the respective shadow is given by \cite{Singh:2017vfr}
\begin{equation}
     R_S = \frac{r_c}{\sqrt{f(r_c)}}
     \label{159}
\end{equation}
where $r_c$ is the solution of (\ref{circ}): the radius of the unstable null circular geodesic defining the photon sphere, given, as we saw, by (\ref{rct}).

The asymptotic flatness of the black hole is crucial for the result (\ref{159}) to be valid, in Einstein gravity and also in the presence of higher derivative corrections \cite{Konoplya:2020bxa}. For the radius of the shadow of AdS black holes there is a correction term, even in the absence of higher derivatives \cite{Das:2019sty}.

Comparing (\ref{omegageo}) with (\ref{159}), we can immediately notice a relation between the shadow radius $R_S$ and the real part of the eikonal quasinormal frequencies associated with test scalar fields, as noticed in \cite{Jusufi:2019ltj,Cuadros-Melgar:2020kqn}:
\begin{equation}
     R_S = \frac{1}{\Omega}.
     \label{160}
\end{equation}
In the absence of higher derivative corrections, this relation extends to all quasinormal frequencies, since their eikonal limits are all identical.

Also crucial for the equality (\ref{159}) to be valid is the fact that photons follow a null geodesic of the background black hole metric. If we were considering black holes in general relativity coupled to nonlinear electrodynamics, as in \cite{Toshmatov:2018tyo, Stuchlik:2019uvf}, photons would rather propagate along the null geodesics of an effective metric characterized by the nonlinearities of the field. This way, the black hole shadow would be governed by the effective geometry directly reflecting the electrodynamic non-linearity, while the quasinormal modes would still be governed by the spacetime geometry, and there would be no relation like (\ref{160}) between them.

Using (\ref{omegat}) and (\ref{160}), we can relate the shadow radius $R_S$ of the Callan-Myers-Perry black hole to the horizon radius $R_H$:
\begin{equation}
    R_S =\sqrt{\frac{2}{d-3}} \left(\frac{d - 1}{2}\right)^{\frac{d-1}{2(d-3)}} \left[1 + \frac{\lambda}{R_H^2}\frac{d-4}{2} \left(1- \left(\frac{2}{d - 1}\right)^{\frac{d-1}{d-3}} \right) \right] R_H.
    \label{220}
\end{equation}

Equation (\ref{220}) can be inverted, in order to obtain $R_H$ expressed in terms of $R_S$. This way, as we mentioned, we can obtain information about the near-horizon geometry from the black hole shadow.

Physical quantities such as the black hole mass and temperature, which are written as functions of $R_H$, can then be expressed in terms of $R_S$ and be estimated, if the shadow radius of a black hole that can be described by the metric (\ref{fr2}) turns to be measured. Other physical quantities such as the quasinormal frequencies can also be expressed as functions of mass or temperature and can, therefore, be estimated knowing $R_S$. Since quasinormal frequencies can also be independently measured, at least in some limits, these estimates can be confronted with experimental results.

%%%%%%%%%%%%%%%%%%%%%%%%%%%%%%%%%%%%%%%%%%%%%%%%%%%%%%%%%%%%%%%%%%%%%%
\section{Conclusions and future directions}
%%%%%%%%%%%%%%%%%%%%%%%%%%%%%%%%%%%%%%%%%%%%%%%%%%%%%%%%%%%%%%%%%%%%%%
\noindent

In this work we computed the quasinormal frequencies in the eikonal limit corresponding to test fields and tensorial gravitational
perturbations for the simplest case of a $d-$dimensional spherically symmetric black hole solution with string corrections given by the Callan-Myers-Perry black hole (\ref{fr2}). We expressed the results in terms of the black hole temperature and we concluded that $\a$ corrections are positive for both real and imaginary parts of the QNM frequencies. Knowing the real part of the frequency corresponding to test fields, we obtained the radius of the black hole shadow.

The most surprising result we obtained was the equality (\ref{llt}) between the imaginary parts of the two different quasinormal frequencies. The proof of the equality (\ref{llt}) we provided in section \ref{equality} is valid for the solution (\ref{fr2}) we worked with, and for the
potential (\ref{vt}) for tensorial perturbations. Both the solution and the potential result from the specific string-theoretical action
(\ref{eef}). One may wonder if a property similar to (\ref{llt}) is valid in general for other theories with higher derivative corrections (such as Lovelock theories), or if it is an intrinsic property of string-theoretical solutions, or eventually just of this solution. We obtained a condition  (\ref{204}) that a generic spherically symmetric black hole solution with higher derivative corrections should obey in order to verify (\ref{llt}).

In Lovelock theories, differently than in string theory, the higher-derivative corrections are not seen as a perturbative expansion. This way, the parameter controlling e.g. the Gauss-Bonnet term may be taken to arbitrary order in the solution to the (nonlinear higher order) field equations, even if it appears only to first order in the lagrangian. In \cite{Konoplya:2017wot} quasinormal modes in the eikonal limit of Gauss-Bonnet $d-$dimensional black holes have been computed for tensorial, vectorial and scalar gravitational perturbations, up to second order in the Gauss-Bonnet parameter. Similarly to our case, no agreement was found for the real parts of the quasinormal modes even to first order in the Gauss-Bonnet parameter. The agreement between the imaginary parts was observed for all types of perturbations (also for vectorial and scalar gravitational perturbations), but only to first order in the Gauss-Bonnet parameter -- not to second order, where it failed. That is probably the reason the agreement just to first order was overlooked in \cite{Konoplya:2017wot}.

It is tempting to question if the agreement we found could be extended, for the same solution, to vectorial and scalar gravitational perturbations. In order to verify such agreement, one would need to determine the potentials corresponding to such perturbations for solutions corresponding to the stringy lagrangian (\ref{eef}). The determination of such potentials, the study of the stability of the Callan-Myers-Perry black hole (\ref{fr2}) under such perturbations and the calculation of the corresponding quasinormal modes are by themselves relevant problems; the eventual verification of an identity like (\ref{llt}) is an interesting extra motivation. Since in string theory $\a$ corrections come in a perturbative expansion, all these studies we propose here could also be extended to black holes with higher order (namely $\a^3$) corrections. One could then figure out if the identity (\ref{llt}) is a more general property in string theory or if is just a special feature of quadratic curvature corrections.

If one could conclude that the identity (\ref{llt}) was indeed a general property, that would mean that the imaginary part of quasinormal frequencies in the eikonal limit would be equal to the principal Lyapunov exponent corresponding to circular null geodesics. If that statement is true, it could
help to clarify the question of the stability of these black holes under these perturbations in different dimensions, a topic with many open problems. This is why we think this subject deserves further study.

%%%%%%%%%%%%%%%%%%%%%%%%%%%%%%%%%%%%%%%%%%%%%%%%%%%%%%%%%%%%%%%%
%%%%%%%%%%%%%%%%%%%%%%%%%%%%%%%%%%%%%%%%%%%%%%%%%%%%%%%%%%%%%%%%
\paragraph{Acknowledgements}
\noindent
%%%%%%%%%%%%%%%%%%%%%%%%%%%%%%%%%%%%%%%%%%%%%%%%%%%%%%%%%%%%%%%%
%%%%%%%%%%%%%%%%%%%%%%%%%%%%%%%%%%%%%%%%%%%%%%%%%%%%%%%%%%%%%%%%
This work has been supported by Funda\c c\~ao para a Ci\^encia e a Tecnologia under contracts IT (UIDB/50008/2020 and UIDP/50008/2020) and project CERN/FIS-PAR/0023/2019.

\end{document}